\documentclass[aps,showpacs,showkeys,preprintnumbers,amsmath,amssymb]
{revtex4}
\usepackage{graphicx}
\begin{document}

\title{Neutron stars with isovector scalar correlations}
\author{B. Liu$^{1,2}$, H. Guo$^{1,3}$, M. Di Toro$^{4}$,
  V. Greco$^{4,5}$}

\affiliation{$^{1}$ Center of Theoretical Nuclear Physics,
National Laboratory of Heavy Ion Accelerator,  Lanzhou 730000,
China}
\affiliation {$^{2}$ Institute of High Energy Physics, Chinese
Academy of Sciences, Beijing 100039, China}
\affiliation{$^{3}$ Department of Technical Physics, Peking
University, Beijing 100871, China}
\affiliation {$^{4}$  Laboratori Nazionali del Sud, Via S. Sofia 44,
I-95123 Catania and
University of Catania, Italy}
\affiliation {$^{5}$ Cyclotron  Institute, Texas A \& M University,
College Station, USA}

\begin{abstract}

Neutron stars with the isovector scalar $\delta$-field are
studied in the framework of the relativistic mean field ($RMF$)
approach in a pure nucleon plus lepton scheme. The $\delta$-field leads to
a larger repulsion in dense neutron-rich matter and to a definite splitting of
proton and neutron effective masses. Both features are influencing the
stability conditions of the neutron stars.
Two parametrizations for the effective nonlinear
Lagrangian density are used to calculate the nuclear equation of
state ($EOS$) and the neutron star properties, and compared to
correlated Dirac-Brueckner results. We conclude that in order to
reproduce reasonable nuclear structure and neutron star properties
within a $RMF$ approach a density dependence of the coupling constants
is required.

\end{abstract}



\maketitle

 A Relativistic Mean Field ($RMF$) approach to nuclear
matter with the coupling to an isovector scalar field, a
virtual $a_{0}(980)$
$\delta$-meson, has been studied for the asymmetric
nuclear matter at low densities, including its
linear response \cite{s1,s2,s3}, and for heavy ion collisions
at intermediate energies where larger density and momentum regions
can be probed \cite{s4,s5,s6}.
In this work we extend the analysis of the contribution of the
$\delta$-field in dense asymmetric matter to the
influence on neutron star properties.
We also want to
test which effective interaction is more appropriate for the description of
dense matter, including the symmetric part. In order to make a
comparison, we use the two parametrizations for the effective
nonlinear Lagrangian density to study the strongly
isospin-asymmetric matter at high density regions.

A Lagrangian density  of the interacting many-particle
system consisting of nucleons, isoscalar (scalar $\sigma$, vector $\omega$),
and isovector (scalar $\delta$, vector $\rho$) mesons is the starting
point of the $RMF$ theory

\begin{widetext}
\begin{eqnarray}\label{eq:1}
{\cal L } &=& \bar{\psi}[i\gamma_{\mu}\partial^{\mu}-(M-
g_{\sigma}\phi -g_{\delta}\vec{\tau}\cdot\vec{\delta})
-g{_\omega}\gamma_\mu\omega^{\mu}-g_\rho\gamma^{\mu}\vec\tau\cdot
\vec{b}_{\mu}]\psi \nonumber \\&&
+\frac{1}{2}(\partial_{\mu}\phi\partial^{\mu}\phi-m_{\sigma}^2\phi^2)
-U(\phi)+\frac{1}{2}m^2_{\omega}\omega_{\mu} \omega^{\mu}
+\frac{1}{2}m^2_{\rho}\vec{b}_{\mu}\cdot\vec{b}^{\mu} \nonumber
\\&&
+\frac{1}{2}(\partial_{\mu}\vec{\delta}\cdot\partial^{\mu}\vec{\delta}
-m_{\delta}^2\vec{\delta^2}) -\frac{1}{4}F_{\mu\nu}F^{\mu\nu}
-\frac{1}{4}\vec{G}_{\mu\nu}\vec{G}^{\mu\nu},
\end{eqnarray}
\end{widetext}

where $\phi$ is the $\phi$-meson field,
$\omega_{\mu}$ is the $\omega$-meson field, $\vec{b}_{\mu}$ is
$\rho$ meson field, $\vec{\delta}$ is the isovector scalar field of the
$\delta$-meson.
$F_{\mu\nu}\equiv\partial_{\mu}\omega_{\nu}-\partial_{\nu}\omega_{\mu}$,
¥$\vec{G}_{\mu\nu}\equiv\partial_{\mu}\vec{b}_{\nu}-\partial_{\nu}\vec{b}_{\mu}$,
and the $U(\phi)$ is a nonlinear potential of $\sigma$ meson :
$U(\phi)=\frac{1}{3}a\phi^{3}+\frac{1}{4}b\phi^{4}$.

The field equations in a mean field approximation ($MFA$) are

\begin{eqnarray}\label{eq:2}
&& (i\gamma_{\mu}\partial^{\mu}-(M- g_{\sigma}\phi
-g_\delta{\tau_3}\delta_3)-g_\omega\gamma^{0}{\omega_0}-g_\rho\gamma^{0}{\tau_3}{b_0})\psi=0,\nonumber \\
&& m_{\sigma}^2\phi+ a{{\phi}^2}+ b{{\phi}^3}=g_\sigma<\bar\psi\psi>=g_\sigma{\rho}_s, \nonumber \\
&& m^2_{\omega}\omega_{0}=g_\omega<\bar\psi{\gamma^0}\psi>=g_\omega\rho,\nonumber \\
&& m^2_{\rho}b_{0}=g_\rho<\bar\psi{\gamma^0}\tau_3\psi>=g_\rho\rho_3,\nonumber \\
&& m^2_{\delta}\delta_3=g_{\delta}<\bar\psi\tau_3\psi>=g_{\delta}\rho_{s3},
\end{eqnarray}

where $\rho_3=\rho_p-\rho_n$ and
$\rho_{s3}=\rho_{sp}-\rho_{sn}$, $\rho$ and $\rho_s$ are
the baryon and the scalar densities, respectively.


Neglecting the derivatives of mesons fields, the energy-momentum tensor in
$MFA$ is given by

\begin{equation}\label{eq:3}
T_{\mu\nu}=i\bar{\psi}\gamma_{\mu}\partial_{\nu}\psi+[\frac{1}{2}
 m_{\sigma}^2\phi^2+U(\phi)+\frac{1}{2}m_{\delta}^2\vec{\delta^2}
-\frac{1}{2}m^2_{\omega}\omega_{\lambda} \omega^{\lambda}
-\frac{1}{2}m^2_{\rho}\vec{b_{\lambda}}\vec{b^{\lambda}}]g_{\mu\nu}.
\end{equation}

The equation of state ($EOS$) for nuclear matter at T=0 is straightforwardly
obtained from the energy-momentum tensor.
The energy density has the form

\begin{equation}\label{eq:4}
\epsilon=<T^{00}>=
\sum_{i=n,p}{2}\int \frac{{\rm d}^3k}{(2\pi)^3}E_{i}^\star(k))
+\frac{1}{2}m_\sigma^2\phi^2+U(\phi)
+\frac{1}{2}m_\omega^2\omega_0^2
+\frac{1}{2}m_{\rho}^2 b_0^2,
+ \frac{1}{2}m_{\delta}^2\delta_3^2,
\end{equation}

and the pressure is

\begin{equation}\label{eq:5}
p =\frac{1}{3}\sum_{i=1}<T^{ii}>=
\sum_{i=n,p} \frac{2}{3}\int \frac{{\rm d}^3k}{(2\pi)^3}
\frac{k^2}{E_{i}^\star(k)}
 -\frac{1}{2}m_\sigma^2\phi^2-U(\phi)+
\frac{1}{2}m_\omega^2\omega_0^2
+\frac{1}{2}m_{\rho}^2{b_0^2}
-\frac{1}{2}m_{\delta}^2\delta_3^2,
\end{equation}

where ${E_i}^\star=\sqrt{k^2+{{M_i}^\star}^2}$, i=p,n. The nucleon
effective masses are, respectively

\begin{equation}\label{eq:6}
{M_p}^\star=M-g_\sigma\phi-g_\delta\delta_3,
\end{equation}

and

\begin{equation}\label{eq:7}
{M_n}^\star=M-g_\sigma\phi+g_\delta\delta_3.
\end{equation}

The nucleon chemical potentials $\mu_i$ are given in terms of the
vector meson mean fields

\begin{equation}\label{eq:8}
\mu_i=\sqrt{k_{F_i}^2+{{M_i}^\star}^2}+ g_\omega\omega_0\mp g_{\rho}b_0\
~~(+~proton, -~neutron),
\end{equation}

where the Fermi momentum $k_{F_i}$ of the nucleon is related to its
density, $k_{F_i}=(3\pi^2 \rho_i)^{1/3}$.

Since we are interested in the effects of the Nuclear Equation of State
we will consider only pure nucleonic (+lepton) neutron
star structures, i.e. without strangeness bearing baryons and even
deconfined quarks, see the recent nice review \cite{s7} and the ref.\cite{s8}.
In particular we will use two models for the neutron star
composition, pure neutron and  $\beta$-stable  matter.
In the latter case we limit
the constituents to be neutrons, protons and electrons.
Then the composition is determined by the request of charge neutrality and
$\beta$-equilibrium.
The $(n,p,e^{-})$ matter is indeed the most important $\beta$-stable
nucleon + lepton matter at low temperature.

The chemical potential equilibrium condition for the $(n,p,e^{-})$ system
can be written as

\begin{equation}\label{eq:9}
\mu_{e} =\mu_{n}-\mu_{p}~.
\end{equation}

\noindent
The charge neutrality condition is

\begin{equation}\label{eq:10}
\rho_{e} =\rho_{p}=X_{p}\rho~,
\end{equation}

\noindent
where $X_{p}=Z/A=\rho_{p}/\rho$ is the proton fraction (asymmetry parameter
$\alpha=1-2X_{p}$), and $\rho$ is the total baryon density.
The electron density $\rho_{e}$ in the ultrarelativistic
limit for non-interacting electrons can be denoted as a function of
its chemical potential

\begin{equation}\label{eq:11}
\rho_{e} =\frac{1}{3\pi^2} \mu_{e}^{3}~,
\end{equation}

\noindent
where $\mu_e=\sqrt{k_{F_{e}}^{2}+m_{e}^{2}}$.
The $X_{p}$ can be obtained by using Eqs.(8), (9), (10) and (11).
The $X_{p}$ is related to the nuclear symmetry energy $E_{sym}$

\begin{equation}\label{eq:12}
3\pi^{2}\rho X_{p} - [4E_{sym}(\rho)(1-2 X_{p})]^{3}=0.
\end{equation}

\noindent
In presence of a coupling to an isovector-scalar $\delta$-meson field,
 the expression for the symmetry energy has a simple transparent
form, see \cite{s2,s3}:

\begin{equation}\label{eq:13}
E_{sym}(\rho)=\frac{1}{6} \frac{k_{F}^{2}}{E_{F}}+
\frac{1}{2}[f_{\rho}-f_{\delta}(\frac{M^{\star}}{E_{F}^{\star}})^{2}]\rho~,
\end{equation}

\noindent
where $M^{\star}=M-g_{\sigma}\phi$ and
${E_F}^\star=\sqrt{k_{F}^2+{M^\star}^2}$.
The $E_{sym}$ and  the $EOS$ for the $\beta$-stable ($npe^-$) matter at
T=0 can be
estimated by using the obtained values of $X_{p}$.
Equilibrium properties of the neutron stars can be finally studied by
solving Tolmann-Oppenheimer-Volkov ($TOV$) equations \cite{s9,s10}
inserting the derived nuclear $EOS$ as an input, \cite{s7}.

The isovector coupling constants, $\rho$-field and $\rho+\delta$ cases,
are fixed from the symmetry energy
at saturation and from Dirac-Brueckner estimations, see the detailed
discussions in refs.\cite{s2,s3}.

In order to make a comparison, two parameter sets for
the isoscalar part are used. The first, $Set~A$, is more suitable at 
high densities where
it appears closer to various Dirac-Brueckner predictions.
 In fact recently this interaction has been used with success to describe
reaction observables in $RMF$-transport simulations of relativistic
heavy ion collisions, where high densities and momenta are reached
\cite{s4,s5,s6}.
The second,
 $Set~B$, is taken from the $NL3$ parametrization \cite{s11}, obtained by
fitting
properties of symmetric nuclear matter at saturation density and
of finite nuclei. In ref.\cite{s5} it has been shown that the good
description of finite nuclei, even exotic, is kept also when the
isovector-scalar channel is included, normally not present in the
$NL3$ Lagrangian.

The coupling constants, $f_{i}\equiv g_{i}^{2}/m_{i}^{2}$,
$i=\sigma, \omega, \rho, \delta$, and the two parameters of the $\sigma$
self-interacting terms : $A\equiv a/g_{\sigma}^{3}$ and $B\equiv
b/g_{\sigma}^{4}$
are reported in Table 1.
The corresponding properties of nuclear matter are listed in Table 2.

\par
\vspace{0.3cm}
\noindent

\begin{center}
{{\large \bf Table 1.}~Parameter sets.}
\par
\vspace{0.5cm}
\noindent

\begin{tabular}{c|c|c|c|c} \hline
Parameter  &\multicolumn{2}{|c|}{$Set~A$}
           &\multicolumn{2}{|c}{$Set~ B$} \\ \cline{2-5}
  &NL$\rho$  &NL$\rho\delta$   &NL$\rho$  &NL$\rho\delta$ \\ \hline
$f_\sigma~(fm^2)$  &10.32924  &10.32924    &15.61225      &15.61225 \\ \hline
$f_\omega~(fm^2)$  &5.42341   &5.42341     &10.40068      &10.40068  \\ \hline
$f_\rho~(fm^2)$    &0.94999   &3.1500      &1.09659       &3.08509 \\ \hline
$f_\delta~(fm^2)$  &0.00      &2.500       &0.00          &2.400    \\ \hline
$A~(fm^{-1})$      &0.03302   &0.03302     &0.00999       &0.00999   \\ \hline
$B$                &-0.00483  &-0.00483    &-0.002669     &-0.002669  \\ \hline
\end{tabular}
\end{center}
\vspace{0.5cm}
\noindent

\begin{center}

{{\large \bf Table 2.}~Saturation properties of nuclear matter.}

\par
\vspace{0.5cm}
\noindent

\begin{tabular}{ c c c } \hline
$Parameter~~sets$    &A      &B  \\ \hline
$\rho_{0}~(fm^{-3})$ &0.16   &0.148  \\ \hline
$E/A ~(MeV)$         &-16.0  &-16.299 \\ \hline
$K~(MeV)$            &240.0  &271.7  \\ \hline
$E_{sym}~(MeV)$      &31.3   &33.7  \\ \hline
$M^{*}/M $           &0.75   &0.60  \\ \hline
\end{tabular}
\end{center}


\par
\vspace{0.3cm}
\noindent

We first use the two parametrizations to calculate the scalar and
the vector potentials, and the binding energy E/A  for symmetric
nuclear matter ($\alpha$=0.0) as a function of baryon density.
The results are presented in Fig.1.

\begin{figure}[hbtp]
\includegraphics[scale=0.4]{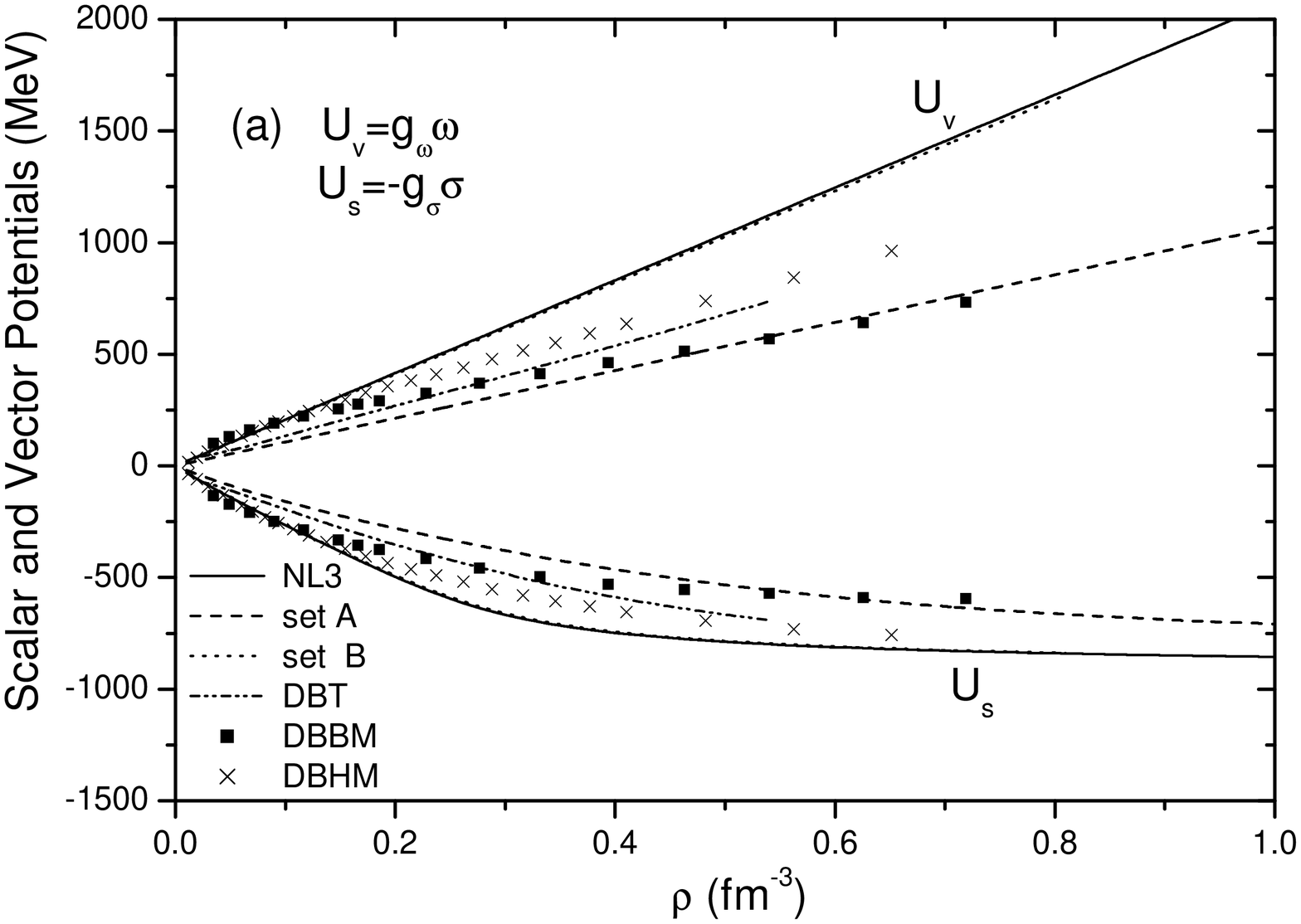}
\vglue -4.0cm
\includegraphics[scale=0.4]{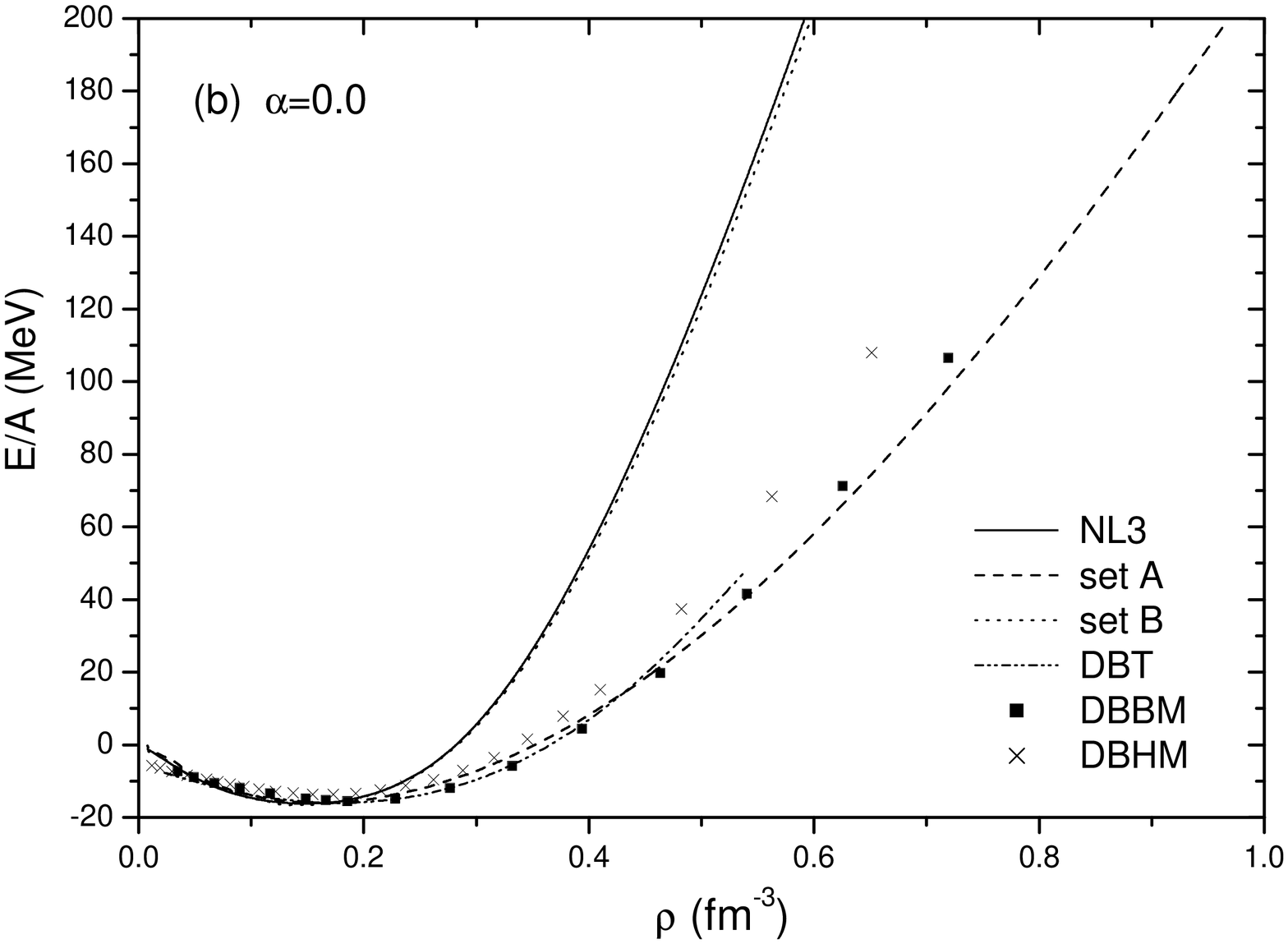}
\vglue -4.8cm
\caption{(a) Scalar and vector potentials vs the
baryon density;(b) binding energy as a function of the baryon
density for symmetric nuclear matter. See text.}
\label{Fig.1}
\end{figure}

The relativistic Dirac-Brueckner-Hartree-Fock ($DBHF$) approach is a
microscopic model describing many-body system with correlations, that has been
extensively used to study the nuclear matter properties
 \cite{s12,s13,s14,s15,s16,s17}.
In order to make a comparison, in Fig.1 we also report the different results
within the $DBHF$ approaches,
the relativistic Dirac-Brueckner calculations by Brockmann and
Machleidt \cite{s12}, denoted as $DBBM$, the Dirac-Brueckner T-matrix
calculations \cite{s13}, denoted as $DBT$, and the
Dirac-Brueckner results by ter Haar and Malfliet \cite{s14}, denoted
as $DBHM$.

From Fig.1(a) we see that the scalar and vector potentials for
the isoscalar channels, given by the
$Set~B$  (i.e. the $NL3$ $\sigma\omega$ couplings) in low density
regions are consistent with the correlated
Dirac-Brueckner
results, while the results given by the $Set~A$ are in better
agreement
in high density regions.

The dotted line in Fig.1(b) denotes
the $EOS$ of symmetric nuclear matter given by the $Set~B$,
in full overlap with the solid line given by the $NL3$ interaction.
It shows a nice agreement with correlated relativistic predictions
at low densities but is clearly too repulsive with increasing density.
At variance
Fig.1(b) also shows that, as expected, the $EOS$ of symmetric nuclear matter
given by the $Set~A$ is more consistent with that given by Dirac-Brueckner
calculations, in particular for the $DBT$ and $DBBM$ estimations,
in high density regions up to about three-four times normal density.
Such very different behaviors at high baryon density, in
connection to the large difference in nucleon effective masses, will strongly
influence the neutron star structure.

Relativistic heavy-ion collisions can provide crucial information
about the $EOS$ of nuclear matter.
The investigation of nuclear $EOS$ at high densities is one of driving force
for study of heavy-ion reactions.
The authors \cite{s15} use different $DBHF$ approaches to study
the collective flow of heavy-ion collisions.
It is shown that the softer $DBT$ choice is in better agreement with
experimental data of relativistic collisions, at least up to a few $AGeV$
beam energies, where densities up to $2.5\rho_0$ can be reached
in the interacting zone.
In general very accurate analyses of
relativistic collisions data favor the predictions of a softer $EOS$
at high densities \cite{s18,s19}.
We remark from  Fig.1(b) that for symmetric matter the $DBT$ is quite close
to our $Set~A$ parametrization, at least up to about $3\rho_{0}$.

\begin{figure}[hbtp]
\begin{center}
\includegraphics[scale=0.4]{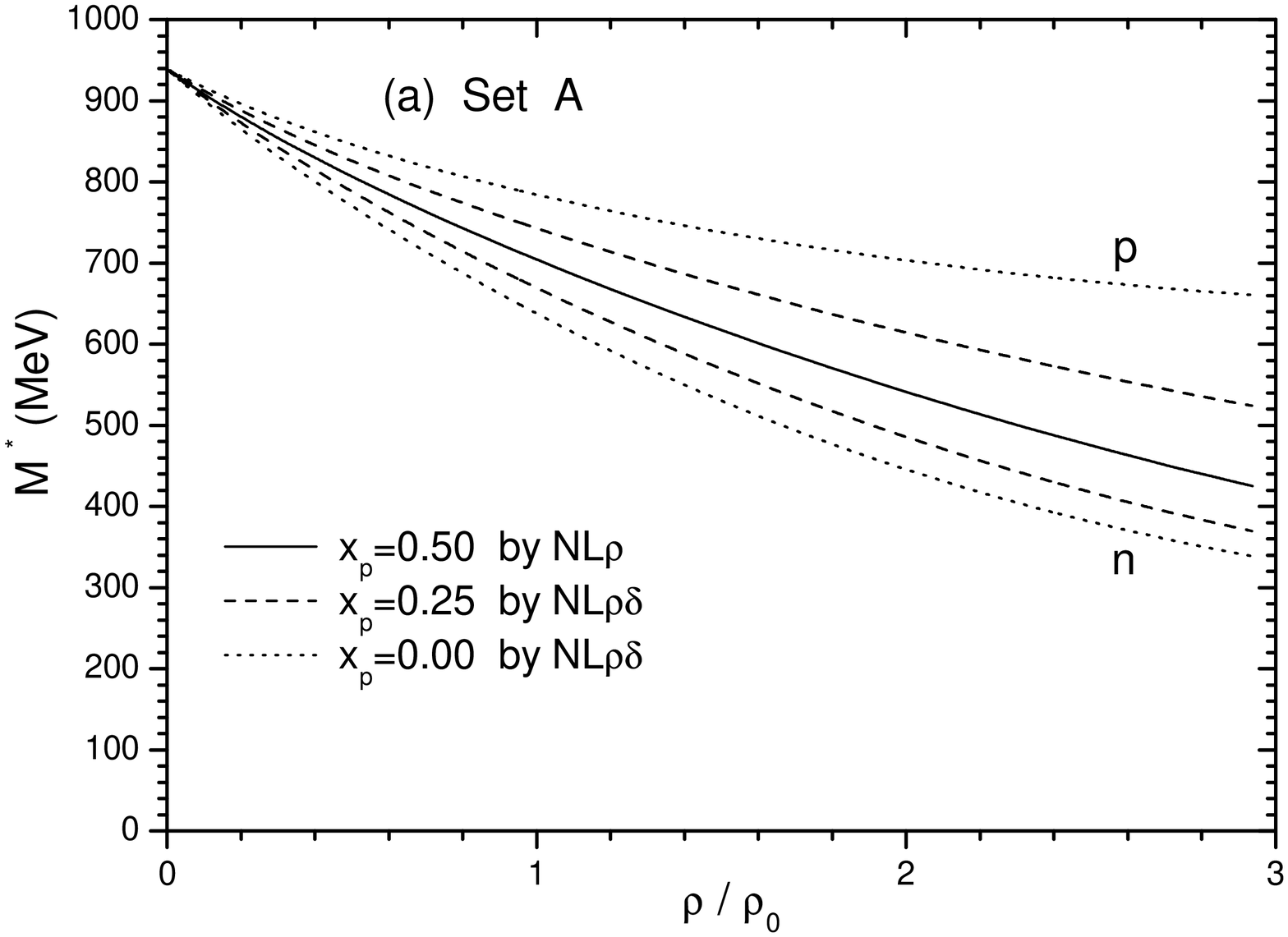}
\vglue -4.4cm
\includegraphics[scale=0.4]{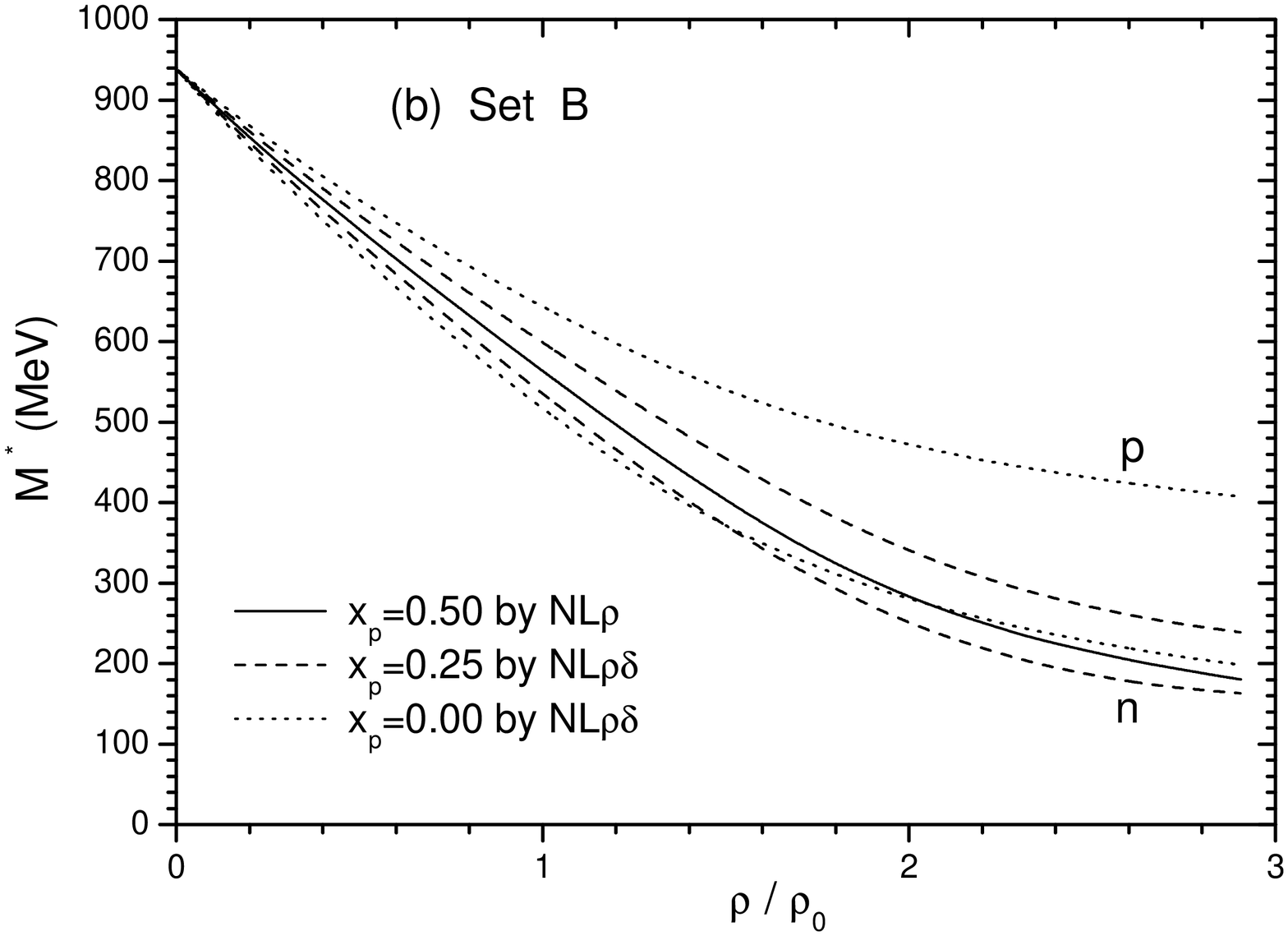}
\vglue -4.6cm
\caption{Neutron and proton effective masses vs.
the baryon density for some values of the proton fraction: (a) by Set A
and (b) by Set B, respectively. See text.}
\label{Fig.2}
\end{center}
\end{figure}

For the isovector channels one can see from Eqs.(6) and (7) that
the presence of the
$\delta$-field leads to proton and neutron effective mass splitting.
In Fig.2 we present the the baryon density dependence of n, p
effective masses for different proton fractions by the two
parameter sets. The solid lines in Fig.2 are the nucleon effective
mass for symmetric nuclear matter ($X_{p}=0.5$).

Fig.2(a) shows
that the proton and the neutron effective masses given by the $Set~A$
decrease slowly with increasing baryon density, at variance
with the $Set~B$ case that presents a much faster decrease,
Fig.2(b).
This main
difference between the two, $A$ and $B$, parametrizations, is actually
coming from the isoscalar part. When coupled to the splitting due to the
isovector $\delta$-field it will have large effects on the n-star
equilibrium features.

\begin{figure}[hbtp]
\begin{center}
\includegraphics[scale=0.4]{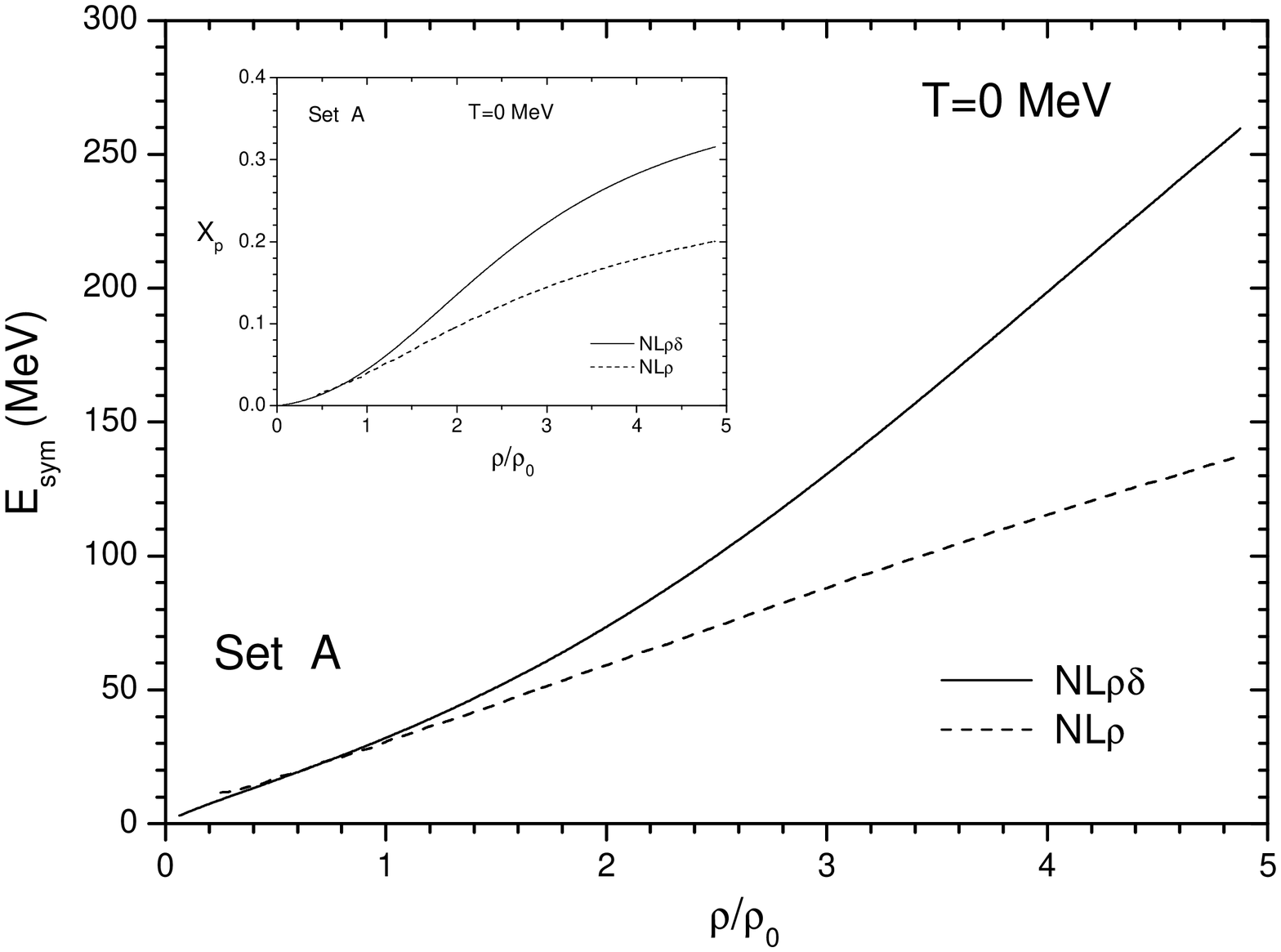}
\vglue -4.6cm
\includegraphics[scale=0.4]{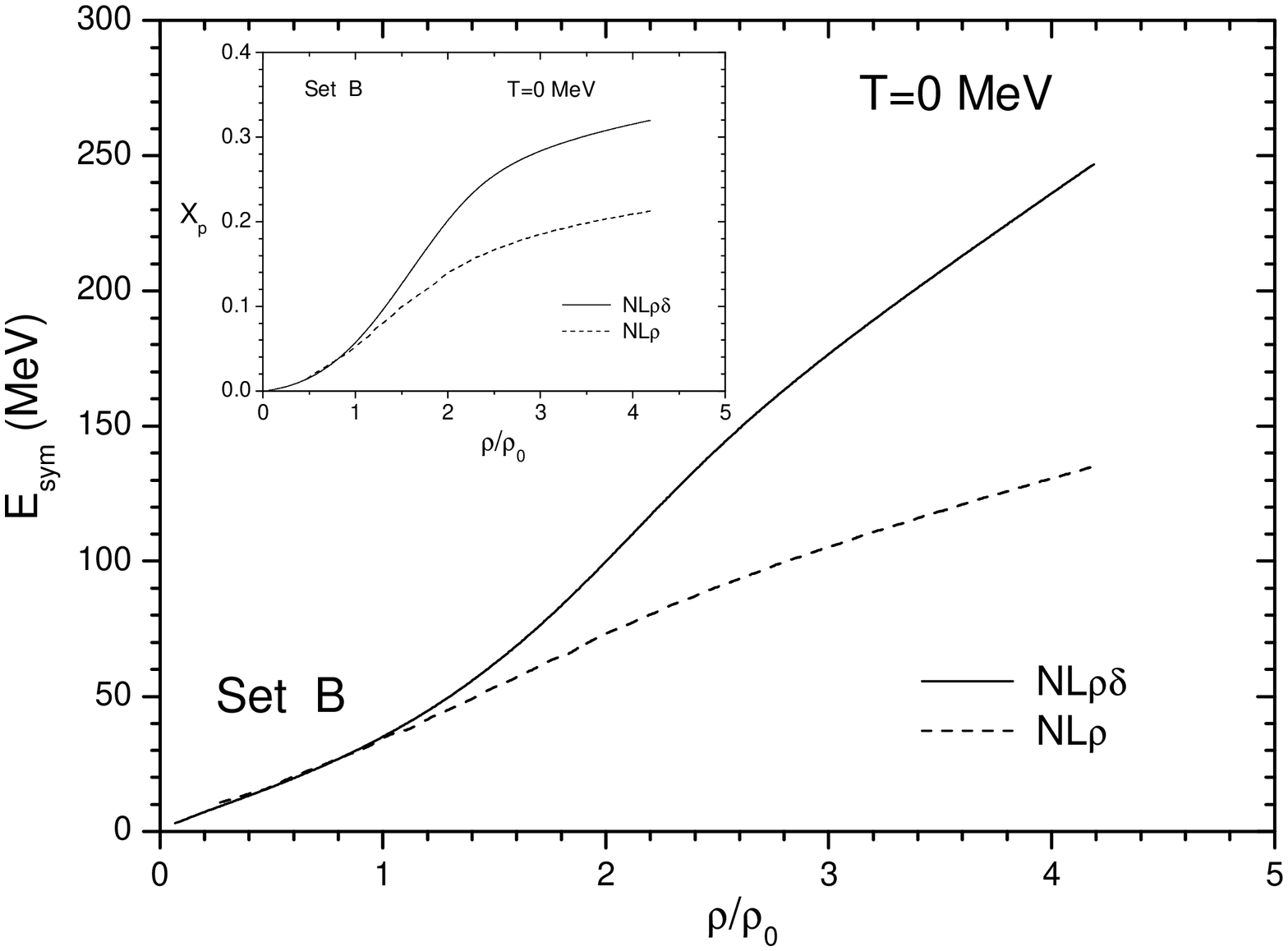}
\vglue -5.0cm
\caption{Symmetry energy vs the baryon density at
T=0 MeV by Set A and Set B, respectively. In the insert the corresponding
proton fraction, see text.}
\label{Fig.3}
\end{center}
\end{figure}

The density dependence of symmetry energy for the two
parameter sets is reported in Fig.3.
For both cases we see a similar behavior of $E_{sym}$ at
sub-saturation densities
for $NL\rho$ and $NL\rho\delta$ models.
With increasing baryon density $\rho$,
however, the differences arising from the presence of the $\delta$-meson
in the isovector channel become more pronounced for both $A$ and $B$,
parametrizations.
This is due to the quenching factor $(M^{\star}/E_{F}^{\star})^{2}$
for the attractive $\delta$ contribution in
Eq. (13) at high density regions, see refs.\cite{s2,s3}.

We note that in spite of
the same isovector coupling constants, at high
density the $Set~B$ gives a larger symmetry energy. This is related to
the larger contribution of the kinetic term in the $r.h.s.$ of Eq.(13)
due to the faster decrease of the effective nucleon mass. This represents
a nice example of how the isovector part of the nuclear $EOS$ can be
influenced by the isoscalar channels due to the Fermi correlations.

In the inserts of Fig.3 we show the corresponding proton fractions $X_{p}$
at $\beta$-equilibrium, Eq. (12). Due to the stiffer nature of the symmetry
energy in the $NL\rho\delta$ cases in both parametrizations we see
a decrease of the $\rho_{Urca}$, i.e. of the baryon density corresponding
to the value $X_{p}=1/9$ that makes possible a direct $Urca$ process,
see \cite{s7}.

\vskip 0.3cm

\begin{figure}[hbtp]
\begin{center}
\includegraphics[scale=0.4]{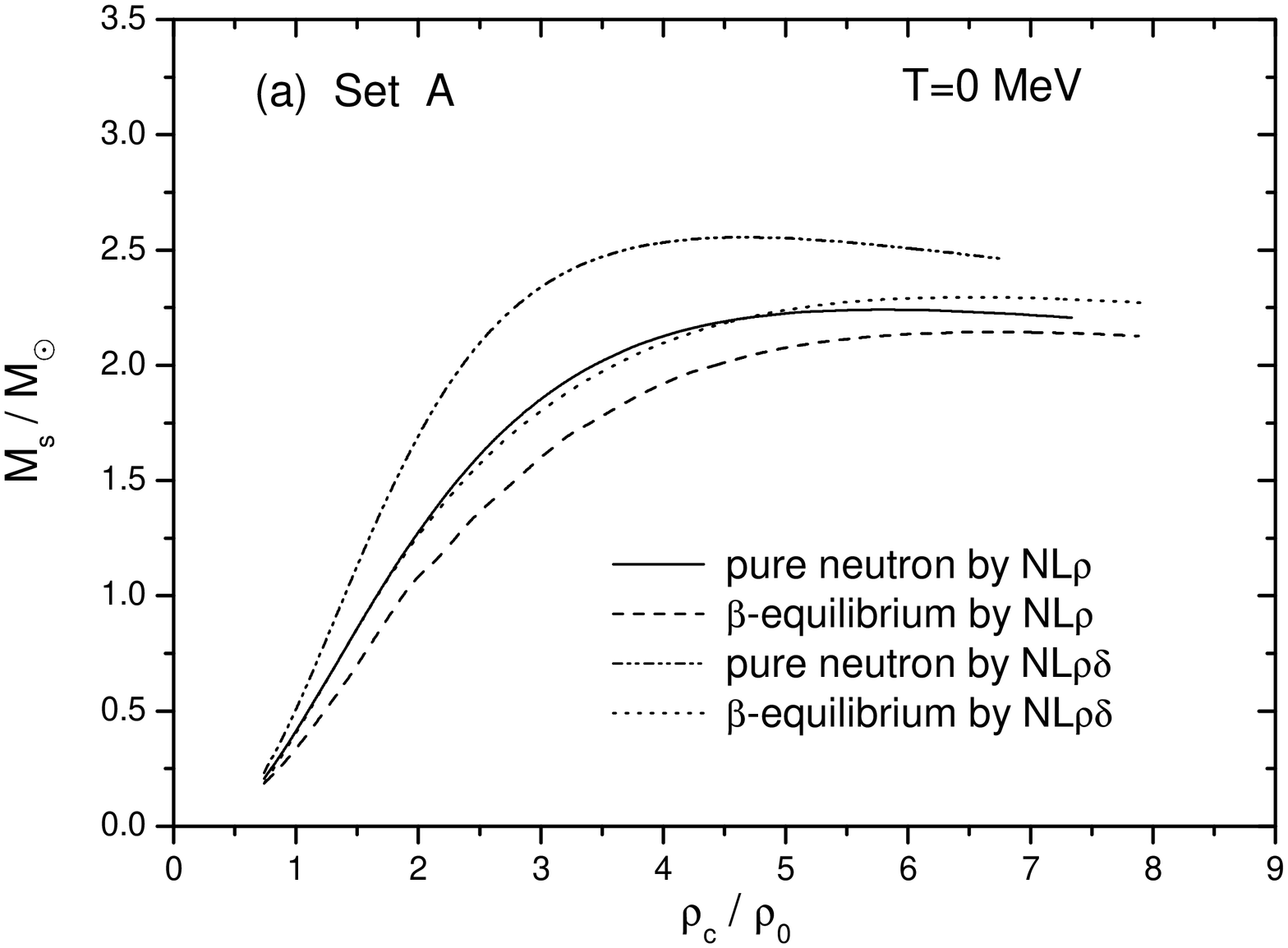}
\vglue -4.2cm
\includegraphics[scale=0.4]{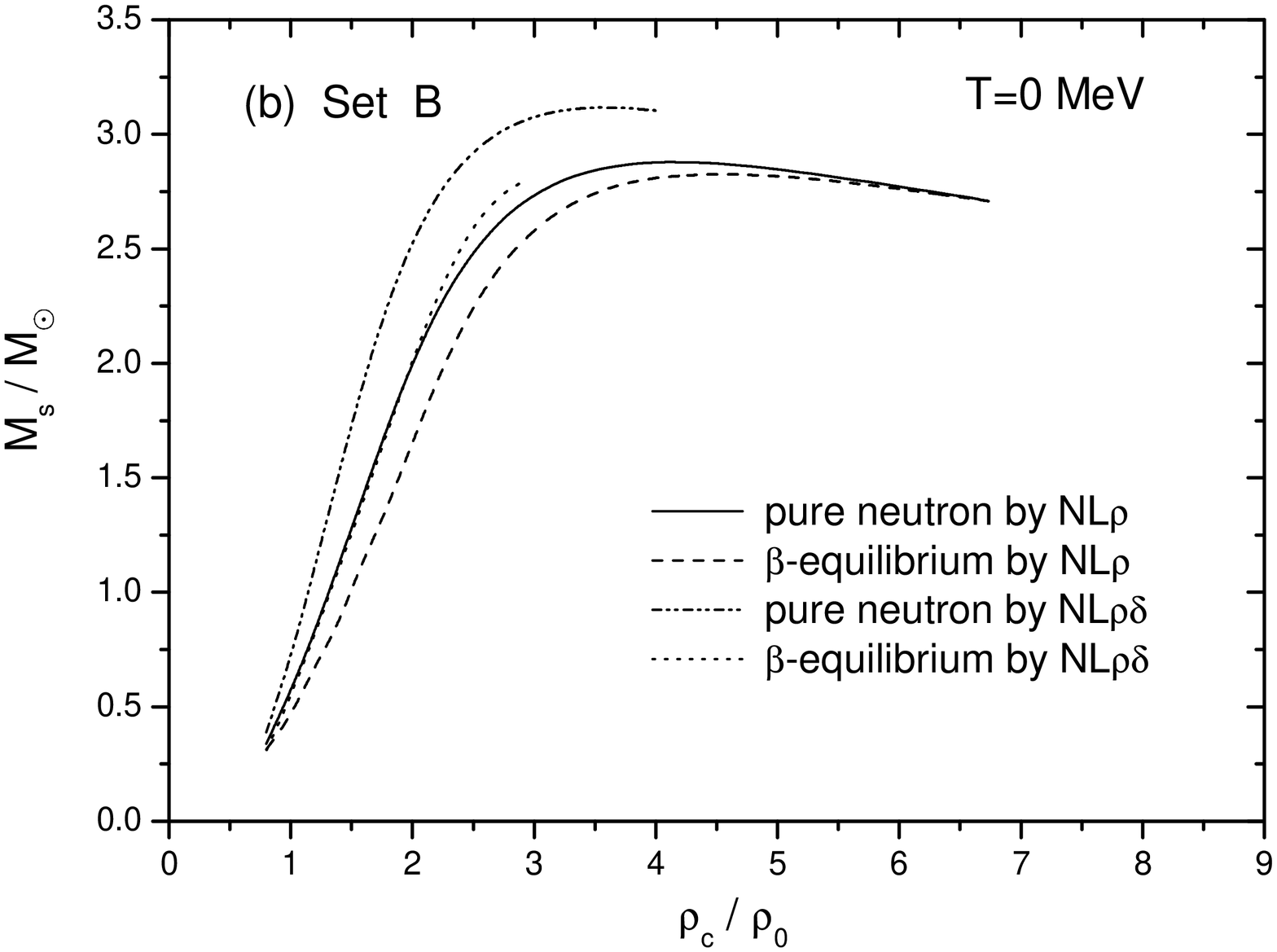}
\vglue -4.7cm
\caption{Mass of the neutron star as a function of the
central density
of the neutron star by Set A and Set B, respectively.}
\label{Fig.4}
\end{center}
\end{figure}

At this point we can derive the predictions for equilibrium properties
of neutron stars just solving the $TOV$ equations \cite{s9,s10}.
The main results are presented in the Figs.4 and 5.

Fig.4  displays the neutron-star mass as a
function of the central density of the star given
by the two parameter sets,  for the two compositions, pure neutron
and the $\beta$-equilibrium  ($npe^-$) matter.

Fig.4(a) and (b) both show that the maximum masses of the
$\beta$-equilibrium star are smaller than in the pure neutron case
 due to the presence of a proton fraction and so of a smaller
symmetry repulsion. Consistently the corresponding central
densities are larger. A related effect is
that the mass of the ($npe^-$) star
given by the $NL\rho\delta$ model decreases more quickly
with increasing density than that given by the $NL\rho$ choice,
 i.e. we have a lower density instability onset.
This is just because the introduction of the $\delta$ coupling increases
the equilibrium proton fraction
at high baryon densities, see the inserts in Fig.3.

The larger stiffness of the symmetry energy in the $NL\rho\delta$
cases, in both parametrizations, can be directly seen in the
fact that the corresponding curves are always above the ones
without the $\delta$ coupling. This implies larger maximum masses
and smaller central densities.
As a consequence we systematically see that the results for the
($npe^-$) composition in the $NL\rho\delta$ models are approximately
equivalent to the ones for the pure neutron matter in the $NL\rho$ choices.

Fig.4(b) presents a qualitatively new feature of the $Set~B$ results:
the lack of solution (maximum mass) in the $\beta$-equilibrium case
for the $NL\rho\delta$ model. The fast decrease of the neutron effective
mass, see Fig.2(b), prevents the chemical potential equilibrium condition
for the ($npe^-$) matter to be satisfied at densities around $3\rho_0$.

\begin{figure}[hbtp]
\begin{center}
\includegraphics[scale=0.4]{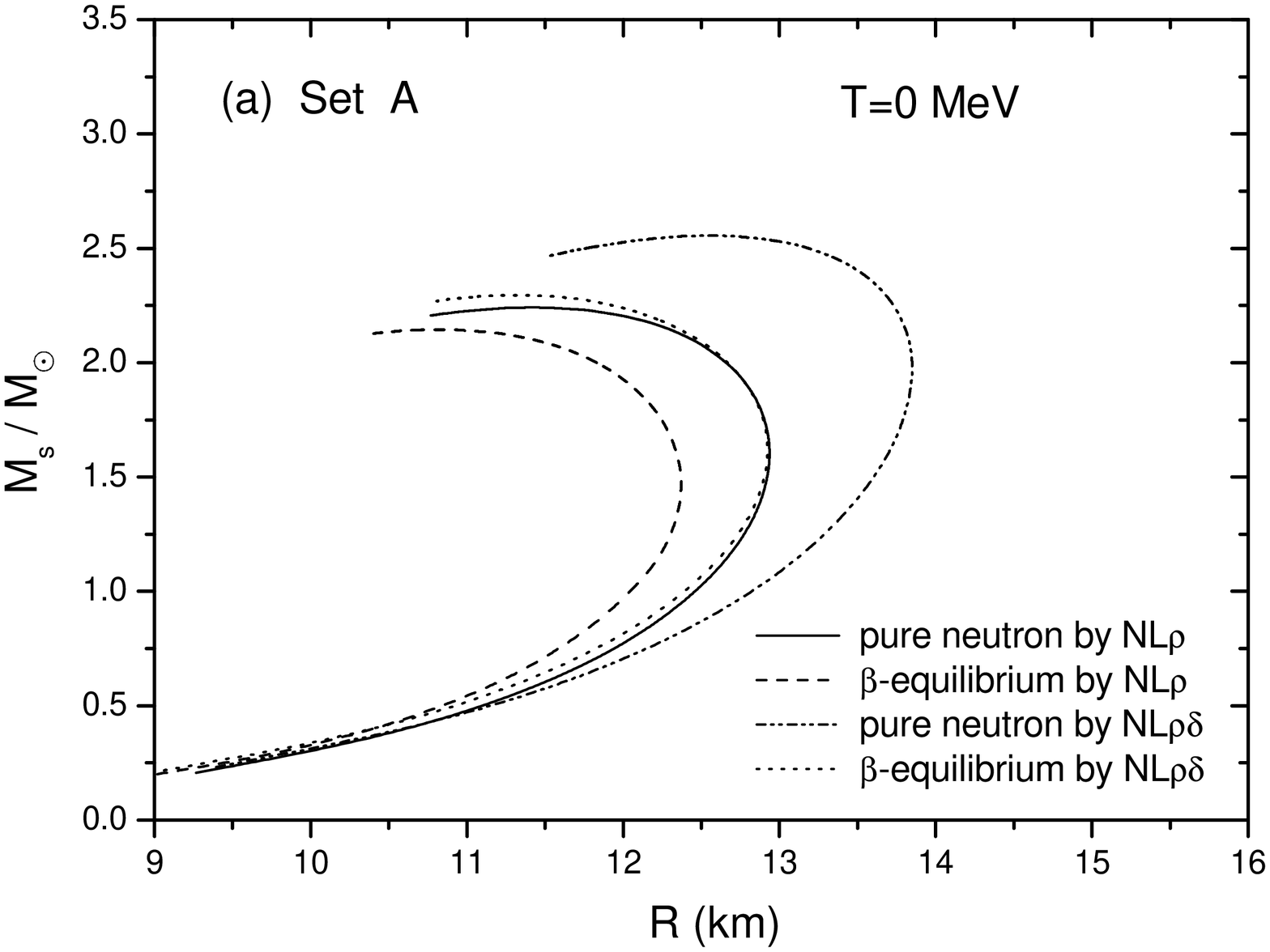}
\vglue -4.4cm
\includegraphics[scale=0.4]{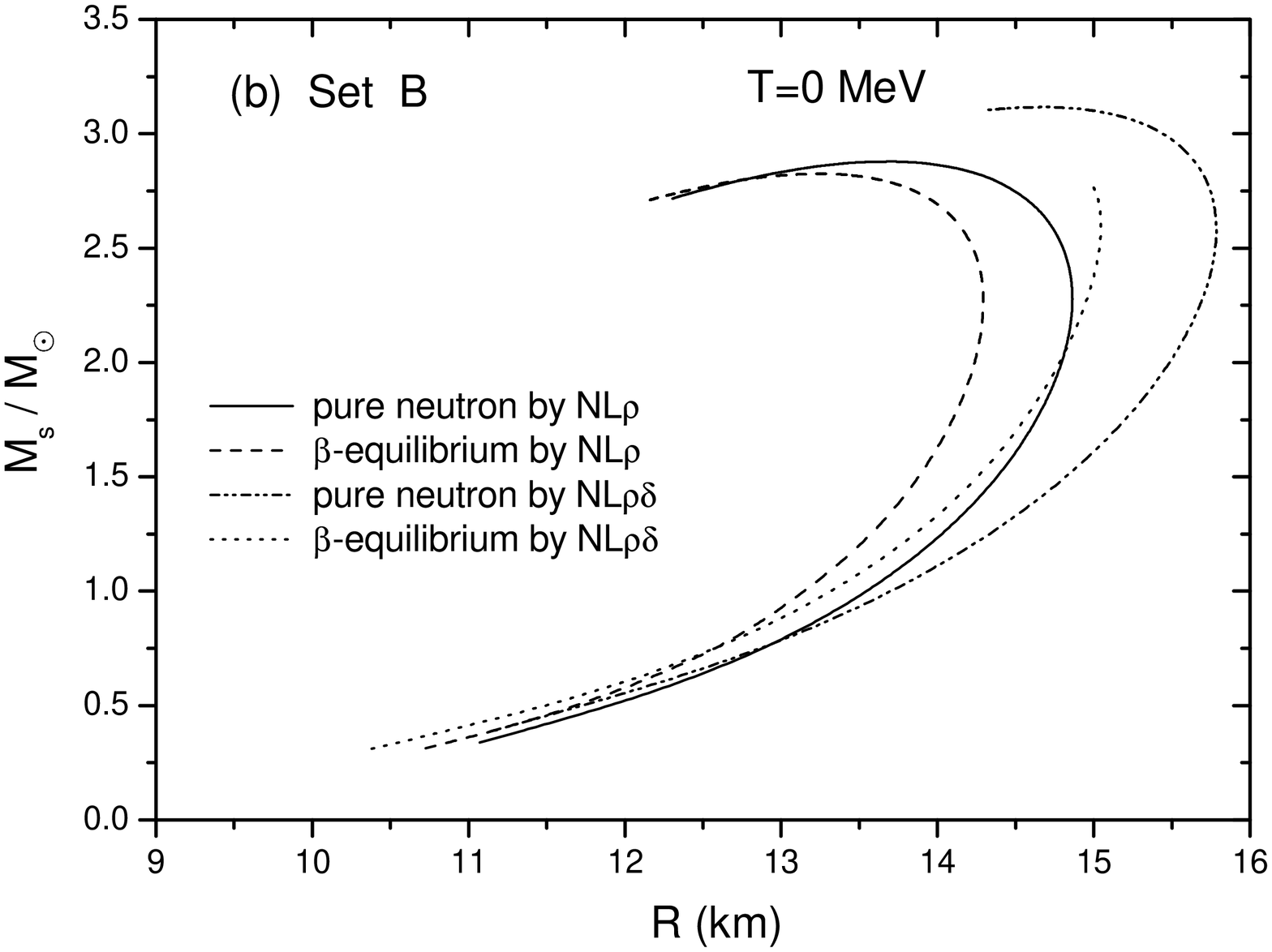}
\vglue -4.6cm
\caption{Mass of the neutron star as a function of
the radius of the neutron star by Set A and Set B, respectively.}
\label{Fig.5}
\end{center}
\end{figure}

Fig.5  reports the correlation between neutron star mass
and radius given by the two
parameter sets, respectively, for the two cases, pure neutron
and ($npe^-$) matter.
Fig.5 shows that the contribution of the $\delta$-field
to the neutron stars at high density regions is quite remarkable.
In particular we note that we systematically have larger masses
and radii and lower central densities, as expected from the larger
symmetry pressure.

All the estimated maximum masses and the corresponding central densities
and radii of the neutron stars are reported in Table 3.

\vspace{0.3cm}

\noindent
\begin{table}
\begin{center}
{{\large \bf Table 3.}~Maximum mass, corresponding radius and
central density of the star given by the two parameter sets.}

\par
\vspace{0.3cm}
\noindent

\begin{tabular}{c|c|c|c|c|c} \hline
 Neutron star   &Properties  &\multicolumn{2}{|c|}{$Set~A$}
                             &\multicolumn{2}{|c}{$Set~ B$} \\ \cline {3-6}
   &  &NL$\rho$  &NL$\rho\delta$   &NL$\rho$  &NL$\rho\delta$ \\ \hline
$pure~ neutron$ &$(M\-{s}/M\-{\odot})_{max}$  &2.24    &2.56    &2.88     &3.12  \\ \cline {2-6}
                &$\rho_{c}/\rho_{0}$          &5.75    &4.65    &4.12     &3.55   \\ \cline {2-6}
                &$R(km)$                      &11.44   &12.56   &13.70    &14.69  \\ \hline
$(npe)~matter$  &$(M\-{s}/M\-{\odot})_{max}$  &2.14    &2.30    &2.83     &        \\ \cline {2-6}
                    &$\rho_{c}/\rho_{0}$      &6.77    &6.49    &4.54     &         \\ \cline {2-6}
                    &$R(km)$                  &10.80   &11.33   &13.25    &          \\ \hline
\end{tabular}
\end{center}
\end{table}

We note that the difference between the two, $A$ and $B$, parametrizations,
in the neutron star predictions,
is largely due to the isoscalar structure of the interactions, see Fig.1.
The $B$ parametrization is much stiffer at high density regions
and this leads to differences on the neutron star masses,
radii and central densities.
The comparison between the results given by the two sets shows
that the set $B$ (i.e. the $NL3$ forces) can be good at low densities
below saturation density and it has serious problems at high densities,
while the set $A$ is a good choice for $EOS$ of nuclear matter at larger
density regions, which is consistent with that pointed out
by refs. \cite{s19,s20}.
In a sense all that just shows that we need density-dependent
$RMF$ parametrizations,
the $B$-type at low densities and $A$-type at high densities.
This has been already emphasized in the work \cite{s21}.
While results of Density Dependent ($DD$) parametrizations and of
the $NL3$ forces agree very well below the saturation density,
the $EOS$ of $DD$-interactions at supra-normal densities shows a much softer
behavior,
similar to $DBHF$ calculations and in better agreement with Heavy Ion
Collision ($HIC$) data,
see also ref. \cite{s5}.

In any case our  calculations show that the $\delta$-field provides
significant contributions
to the neutron star structure for the stiffness of the symmetry energy
and the neutron/proton mass splitting in high density regions.
The proton fraction in the $\beta$-equilibrium matter is much larger than that
in the no $\delta$-field case.
Moreover we can see from Fig.4(b) and Fig.5(b) that we do not have solutions
in the ($npe^-$) case for the set $B$ with $NL\rho\delta$ isovector
interaction.
This represents a quite dramatic effect of the splitting of neutron/proton
effective masses at high densities, on top of the larger nucleon mass
decrease of the $B$
parametrizations, see Fig.2.
The neutron chemical potential is then not able to satisfy
the $\beta$-equilibrium conditions of the ($npe^-$) matter.
The contribution of the $\delta$-field for strongly
isospin-asymmetric dense matter is important and it cannot be neglected.

With reference to neutron star properties we study the $EOS$ for dense
asymmetric matter in the $RMF$ frame with
two different parameter sets for the Lagrangian density. The set $B$ is
close to the $NL3$
parametrization which has been proposed to describe finite nuclei
properties. Since the proton and the neutron effective
masses, especially the neutron effective mass,
decrease quickly with increasing baryon density, the set $B$ cannot
provide the $EOS$ needed in high density regions for the ($npe^-$) star.
This means that in general the $B$ parametrization seems to have serious
problems at
high densities, as already remarked from $HIC$ studies \cite{s19}.
Though the $B$ parametrization can be good for the $EOS$ at low densities,
below and around saturation, it is too stiff in high density
regions, in particular there is no solution for the ($npe^-$) case
with the $\delta$-field. So the set $B$ is not suitable for the case
of dense  matter. We require a softer $EOS$ at high
densities, and indeed the softer Dirac-Brueckner predictions, in
particular the $DBT$ one, are
in better agreement with relativistic collisions data.
 Our $A$  parametrization, including the $\delta$-isovector-scalar field,
is quite close to the $DBT$ and it has been shown to lead to
good predictions in transport simulations for heavy ion collisions at
intermediate energies \cite{s4,s5,s6}.
It appears then quite appropriate for the nucleonic part of the approach
to neutron
star properties. In this respect we note that quite extended neutron star
structure calculations
have been recently performed just using our Set A Lagrangian, 
\cite{MenezesPRC70}.

\section*{Acknowledgments}

This project is supported by the National Natural Science Foundation of China
under Gran No. 10275002, the INFN of Italy, and the Major State Basic Research
Developing Program with grant No. G2000077400.

\vspace{0.5cm}


\end{document}